\documentclass{sig-alternate-05-2015}

\usepackage[numbers,sort]{natbib}
\usepackage{microtype}
\usepackage{mathptmx}
\usepackage{amssymb}
\usepackage{multirow}
\usepackage{amsmath}
\usepackage{graphicx}
\usepackage[table,xcdraw]{xcolor}
\usepackage{chngpage}
\usepackage[caption=false]{subfig}
\usepackage[bottom]{footmisc}
\usepackage{fmtcount}
\usepackage{stfloats}
\usepackage{balance}

\newcommand{\para}[1]{\smallskip\noindent\textbf{#1}}

\makeatletter
\def\@copyrightspace{\relax}
\makeatother

\begin{document}

\title{Discovering and Characterizing Mobility Patterns in\\ Urban Spaces: A Study of Manhattan Taxi Data\titlenote{Please cite the WWW'16 version of this paper.}}

\numberofauthors{2}

\author{
\alignauthor
Lisette Esp{\'i}n-Noboa\\
       \affaddr{GESIS - Leibniz Institute for the Social Sciences}\\
       \email{Lisette.Espin@gesis.org}\\
\alignauthor
Florian Lemmerich\\
       \affaddr{GESIS - Leibniz Institute for the Social Sciences}\\
       \email{Florian.Lemmerich@gesis.org}\\
\and
\alignauthor 
Philipp Singer\\
       \affaddr{GESIS - Leibniz Institute for the Social Sciences}\\
       \email{Philipp.Singer@gesis.org}\\
\and  
\alignauthor 
Markus Strohmaier\\
       \affaddr{GESIS - Leibniz Institute for the Social Sciences \& University of Koblenz-Landau}\\
       \email{Markus.Strohmaier@gesis.org}\\
}

\maketitle

\begin{abstract}
Nowadays, human movement in urban spaces can be traced digitally in many cases.
It can be observed that movement patterns are not constant, but vary across time and space.
In this work, we characterize such spatio-temporal patterns with an innovative combination of two separate approaches that have been utilized for studying human mobility in the past.
First, by using non-negative tensor factorization (NTF), we are able to cluster human behavior based on spatio-temporal dimensions. 
Second, for understanding these clusters, we propose to use HypTrails, a Bayesian approach for expressing and comparing hypotheses about human trails. 
To formalize hypotheses we utilize data that is publicly available on the Web, namely Foursquare data and census data provided by an open data platform.
By applying this combination of approaches to taxi data in Manhattan, we can 
discover and characterize different patterns in human mobility that cannot be identified in a collective analysis.
As one example, we can find a group of taxi rides that end at locations with a high number of party venues (according to Foursquare) on weekend nights.
Overall, our work demonstrates that human mobility is not one-dimensional but rather contains different facets both in time and space which we explain by utilizing online data.
The findings of this paper argue for a more fine-grained analysis of human mobility in order to make more informed decisions for e.g., enhancing urban structures, tailored traffic control and location-based recommender systems.

\end{abstract}


\section{Introduction}
Human mobility can be studied from several perspectives utilizing different kinds of data from the online (e.g., Twitter or Foursquare) and the offline (e.g., taxi rides or bike trips) world. 
A large body of work has focused on identifying general mechanisms that guide and explain human mobility behavior on an individual \cite{gonzalez2008understanding,song2010limits} or collective level \cite{noulas2012tale,becker2015}. 
For example, previous research has shown that human mobility is highly predictable \cite{song2010limits} and shows temporal and spatial regularity \cite{gonzalez2008understanding}. At the same time, spatio-temporal heterogeneity exists, as, for example, discussed in previous studies \cite{liu2013revealing, chen2010study}.
For instance, daily routines such as going from home to work (space) in the morning (time) and from work to home in the evening can be observed.
This argues for a more fine-grained analysis, that goes beyond the universal (mobility) patterns which tend to ignore several aspects of human mobility such as time, weather, race of people or means of transportation. Towards that end, we propose to 
\emph{discover and characterize mobility patterns} in human behavior in urban space.

\para{Material and approach.} 
In this work, we expand existing research~\cite{liu2013revealing, chen2010study} on studying human mobility with a case study using taxi data of Manhattan.
For identifying behavioral differences in terms of time and space, previous research~\cite{peng2012collective,takeuchi2013non} has suggested to
utilize tensor decomposition~\cite{cichocki2009nonnegative, hao2014nonnegative}.
However, interpreting results from tensor decomposition can be difficult and has been based on personal intuitions in previous approaches.
On the other hand, recent research \cite{singer2015hyptrails,becker2015} proposed methods that allow to understand human sequences by comparing hypotheses about the production of mobility trails at interest. Yet, this approach is limited in the sense that it can only explain global behavior without being able to provide more detailed insights. To circumvent these limitations, we propose a unique and original combination of tensor decomposition and HypTrails to understand human behavior on a spatio-temporal level. In particular, we first utilize \emph{non-negative tensor factorization} (NTF)~\cite{cichocki2009nonnegative, hao2014nonnegative} for automatically identifying clusters of mobility behavior. Second, for 
characterizing these clusters, we utilize \emph{HypTrails}~\cite{singer2015hyptrails}---a Bayesian approach for expressing and comparing hypotheses, i.e., transitional assumptions, about human trails.

\para{Findings and contributions.}
In summary, our main contributions are three-fold: First, we present an innovative combination of two methodologies, i.e., non-negative tensor factorization and HypTrails, in order to characterize heterogeneous human mobility behavior. Second, we incorporate existing human mobility patterns into a hypothesis-based schema built upon human beliefs.
Third, we demonstrate the benefits of online data in our attempt to explain human behavior on a spatio-temporal level. As one example, we can discover a group of taxi rides that have drop-off locations with a high number of party venues (according to Foursquare) on weekend nights. 
Results of this study could improve e.g., planning of future events or reconstructions, traffic control, location-based recommender systems to enable public transportation companies to adapt their capacities based on the demand at certain hours or areas.
\section{Datasets}
\label{sec:datasets}
We focus on Manhattan, one of the most densely populated areas in the world,
since it has a huge amount of governmental and private data publicly available on the Web. 
In detail, we study \emph{taxi ride} data in this work. While this data represents a prominent representative of human mobility, our methodology presented in this paper can be applied to other kinds of mobility data.
We represent human mobility as \emph{user trails} which can be seen as single \emph{transitions} between pick-up and drop-off locations of taxi rides. 
In order to construct a rich set of hypotheses for explaining the movement of users, we additionally retrieved information on local venues (e.g., parks, schools, churches) from Foursquare and public census data (i.e., demographics, land-use and socio-economics).
Next, we shortly explain the various datasets in greater detail.

\para{Taxi rides. }
We use publicly available data\footnote{\scriptsize{\url{http://www.andresmh.com/nyctaxitrips/}}} of $173,179,759$ NYC Taxi rides in 2013.
It consists of anonymized records registering when and where a taxi ride started and ended and various features such as the number of passengers or the total fare of the trip.
From all records, we removed taxi rides outside the area (polygon) of Manhattan and some inconsistencies such as records with $\text{trip\_distance} \leq 0$, $\text{trip\_time\_in\_secs} \leq 0$ and $\text{passenger\_count} \leq 0$; our final dataset consists of $143,064,684$ rides.
A description of all attributes and datatypes can be found in the \emph{TLC Taxi Data - API Documentation}\footnote{\scriptsize{\url{https://dev.socrata.com/foundry/data.cityofnewyork.us/gkne-dk5s}}}.

\para{Census data.} 
Instead of working directly with longitude and latitude data, our methodological approach (see Section~\ref{methodology}) operates on a discrete tract state space. 
Tracts are small subdivisions of a county that~\emph{provide a stable set of geographic units for the presentation of statistical data}\footnote{\scriptsize{\url{https://www.census.gov/geo/reference/gtc/gtc_ct.html}}}. We extracted the specifications of all $288$ tracts in Manhattan from the \emph{NYC Planning portal}\footnote{\scriptsize{\url{http://www.nyc.gov/dcp}}}
as a \emph{shape file} from the $2010$ census. 
Having this information, we mapped the GPS coordinates of all taxi rides to one of the $288$ tracts.
Besides the area and location of tracts, we queried the \emph{NYC OpenData}\footnote{\scriptsize{\url{https://nycopendata.socrata.com/}}}
and the \emph{American FactFinder}\footnote{\scriptsize{\url{http://factfinder.census.gov/faces/nav/jsf/pages/index.xhtml}}} databases for accessing relevant online data such as census and land-use. Moreover, we calculated the overlap between landuse-types such as residential and commercial zones and each tract to obtain information on a tract level.

\para{Foursquare venues.} 
According to previous works, human mobility can be explained by the transition of visited places~\cite{peng2012collective} or by intervening opportunities~\cite{stouffer1940intervening, noulas2012tale}. For studying these and similar theories, we gathered information about physical places such as churches, houses and parks situated in Manhattan by querying the Foursquare Search API\footnote{\scriptsize{\url{https://developer.foursquare.com/docs/venues/search}}} to extract places in each tract for $10$ different categories\footnote{\scriptsize{\url{https://developer.foursquare.com/categorytree}}} (e.g., Residence, Work Place or Nightlife Spot). Overall, we collected $153,694$ unique places within Manhattan. Since every venue is described (besides others attributes) by a GPS location, they all were mapped to their respective tract.

\para{Centroids.}
Typically, popular places are most likely to be visited at any time. For this reason, we considered three candidate places as centroids to study whether people visit them or not. Table~\ref{tab:centroids}, shows the geographical coordinates and respective tracts of the center of Manhattan (approximated as the geographical center of Manhattan), the iconic \emph{Flatiron building}
and well-known \emph{Times Square}.

\begin{table}[t]
\vspace{-1em}
\centering
\caption{\textbf{Centroids of places of interest}: This table lists geo-coordinates and corresponding tracts of the geographical center of Manhattan, the Flatiron building and Times Square.}
\label{tab:centroids}
\begin{tabular}{|c|c|c|}
 \hline
 \textbf{Place} & \textbf{lat, lon} & \textbf{Tract ID} \\ \hline
 Manhattan (geo-center) & 40.79090, -73.96640 & 018100 \\ \hline
 Flatiron Building & 40.74111, -73.98972 & 005600 \\ \hline
 Times Square & 40.75773, -73.98570 & 011900 \\ \hline
\end{tabular}
\end{table}

\begin{table*}[t!] 
	\centering
	\caption{\textbf{Universal Theories}: List of 10 universal theories on human mobility. Following the Markov chain assumption, indexes $i$ and $j$ refer to start and end states respectively. Functions $dist(.,.)$, $venues(.)$, and $checkins(.)$ denote the geographical distance between two states, all Foursquare venues in a given state, and the number of check-ins in Foursquare from a given venue, respectively. Vectors $A$ and $B$ represent categorical information of the start and end state, respectively (e.g., race, poverty, land-use).}
	\label{universal-hyp}
	\begin{tabular}{|c|l|c|l|}
		\hline
		\multicolumn{1}{|c|}{\textbf{\#}} & \multicolumn{1}{c|}{\textbf{Mobility Theory}} & \multicolumn{1}{c|}{\textbf{Formula for $q_{i,j}$}} & \multicolumn{1}{c|}{\textbf{Description}} \\ \hline
		\multirow{3}{*}{1} & \multirow{3}{*}{Uniform} & \multirow{3}{*}{$1$} & \multirow{3}{*}{\parbox{8.8cm}{The next location is chosen randomly, independently of the current location.}} \\ 
		& & & \\
		& & & \\ \cline{1-4}
		\multirow{2}{*}{2} & \multirow{2}{*}{Inverse Geographic Distance} & \multirow{2}{*}{$\frac{1}{dist(i,j)}$} & \multirow{2}{*}{\parbox{8.8cm}{The closer the target location, the more likely it is to be visited, see~\cite{ivis2006calculating}.}} \\ 
		& & & \\ \cline{1-4}
		\multirow{5}{*}{3} & \multirow{5}{*}{\parbox{3cm}{Gaussian Distribution}}  & \multirow{5}{*}{$\frac{1}{\sigma \sqrt{2 \pi}}~e^{-\frac{dist(i,j)^2}{2\sigma^2}}$}  & \multirow{5}{*}{\parbox{8.8cm}{The probability of visiting a place $j$ is given by a two-dimensional Gaussian distribution with standard deviation $\sigma$; the smaller the $\sigma$ around the place $i$, the closer to the location $i$. $i$ can either be the current location or a fixed landmark in the city, see~\cite{becker2015}.}} \\
		& & & \\
		& & & \\
		& & & \\
		& & & \\ \cline{1-4}
		\multirow{2}{*}{4} & \multirow{2}{*}{Density}  & \multirow{2}{*}{$|(venues(j)|$}  & \multirow{2}{*}{\parbox{8.8cm}{People get attracted by the amount of venues at destination place.}} \\
		& & & \\ \cline{1-4}
		\multirow{2}{*}{5} & \multirow{2}{*}{Popularity} & \multirow{2}{*}{$\sum\nolimits_{V \in venues(j)} checkins(V)$}  &  \multirow{2}{*}{\parbox{8.8cm}{Venues with more number of check-ins attract more people, see~\cite{noulas2011empirical}.}} \\ 
		& & & \\ \cline{1-4}
		\multirow{3}{*}{6} & \multirow{3}{*}{\parbox{3cm}{Gravitational Mass}} & \multirow{3}{*}{$\frac{venues(i) . venues(j)}{dist(i,j)}$}  & \multirow{3}{*}{\parbox{8.8cm}{People get attracted by the product of forces from the current and close target locations, see~\cite{zipf1946p1}.}} \\
		& & & \\
		& & & \\ \cline{1-4}
		\multirow{3}{*}{7} & \multirow{3}{*}{\parbox{3cm}{Gravitational Target}} & \multirow{3}{*}{$\frac{venues(j)}{dist(i,j)}$}  & \multirow{3}{*}{\parbox{8.8cm}{People get attracted \emph{only} by a force in a target location which is close to the current place. This is a variant of (6)}} \\
		& & & \\
		& & & \\ \cline{1-4}
		\multirow{3}{*}{8} & \multirow{3}{*}{Rank Distance}  & \multirow{3}{*}{$\frac{1}{|\{w: dist(i,w)<dist(i,j)\}|}$}  & \multirow{3}{*}{\parbox{8.8cm}{The probability of moving to a place increases if there are few venues \emph{w} around the current location, see~\cite{noulas2012tale}.}}  \\
		& & & \\
		& & & \\ \cline{1-4}
		\multirow{4}{*}{9} & \multirow{4}{*}{\parbox{3.5cm}{Intervening Opportunities}}  & \multirow{4}{*}{$\frac{|\{w_1: dist(i,w_1)=dist(i,j)\}|}{|\{w_2: dist(i,w_2)<dist(i,j)\}|}$} & \multirow{4}{*}{\parbox{8.8cm}{The number of persons going a given distance is directly proportional to the number of opportunities at that distance and inversely proportional to the number of intervening opportunities, see~\cite{stouffer1940intervening}.}} \\ 
		& & & \\
		& & & \\
		& & & \\ \cline{1-4}
		\multirow{2}{*}{10} & \multirow{2}{*}{Cosine Similarity}  & \multirow{2}{*}{$\frac{A.B}{||A|| ||B||}$} & \multirow{2}{*}{\parbox{8.8cm}{Measures how similar the current and next location are, see~\cite{tan2006introduction}.}} \\
		& & & \\ \cline{1-4}
	\end{tabular}
	\vspace{-1em}
\end{table*}

\section{Methodology}
\label{methodology}

 The goal of this work is to discover and characterize mobility patterns in taxi data, in order to better understand 
 people's travel behavior from one place to another within the city. 
To that end, we propose an innovative combination of two methodologies.
First, we suggest to use \textit{non-negative tensor factorization} (NTF)~\cite{cichocki2009nonnegative} for automatically clustering human mobility behavior. Research has shown that NTF can detect latent features of human mobility in different dimensions such as space and time, cf. for example~\cite{takeuchi2013non}. Second, for intuitively characterize these clusters, we utilize HypTrails~\cite{singer2015hyptrails}---a Bayesian approach for expressing and comparing hypotheses about human trails. We outline in the following both methodological components of this work, but refer to the original publications for details.

\para{Clustering mobility patterns.}
For clustering the data, we utilize NTF which decomposes a given $n$-way tensor $\underline{\mathbf{X}}$ into $n$ components (matrices) that approximate the original tensor when multiplied with each other.
Each matrix contains information on $r$ factors (clusters). In this paper, we define a three-way tensor of taxi rides whose dimensions capture human transitions from one place to another at a certain time: pick-up tract, drop-off tract and pickup time (hour of week) respectively; thus, clustering in terms of both space and time.
Each element of every component determines the scale of mobility flow (weight) with respect to the corresponding factor. In other words, the higher the weight, the more dominant that instance is in that cluster.
Similar to other clustering methods, defining the number of clusters is arbitrary. However, there exist some guidelines to determine a good value of $r$, see e.g., ~\cite{gauvin2014detecting}. In this work, we are not focused on finding the most appropriate number of behavioral components, but being able to characterize different behaviors, which will be detailed in the following section.

\para{Characterizing clustered mobility patterns.}
\label{met-pat-exp}
For characterizing the clustered human mobility behavior, we utilize
HypTrails~\cite{singer2015hyptrails}, a Bayesian approach for expressing and comparing hypotheses about human trails.
Technically, HypTrails models the data with first-order Markov chain models where the state space contains all $288$ tracts of Manhattan.
Fundamentally, hypotheses are represented as matrices $Q$ expressing beliefs in Markov transitions; Section~\ref{hypotheses} describes the hypotheses about human mobility used in this work. Elements $q_{i,j}$ indicate the belief in the corresponding transition probability between states $i$ and $j$; higher values refer to higher belief. The main idea of HypTrails  is to incorporate these hypotheses as Dirichlet priors into the Bayesian inference process. 
HypTrails automatically elicits these priors from expressed hypotheses; an additional parameter $k$ (weighting factor) needs to be provided to steer the overall belief in a hypothesis.
For then comparing the relative plausibility of hypotheses, HypTrails utilizes the \emph{marginal likelihood} (\emph{evidence}) of the Bayesian framework which describes the probability of a hypothesis given the data. 
We can infer the partial ordering based on the plausibility of a given set of hypotheses 
by ranking their evidences from the largest to the smallest for a specific value of $k$.

\section{Hypotheses}
\label{hypotheses}

In this section, we describe our different hypotheses at interest that are used to explain the individual clusters with HypTrails.
These are expressed as hypothesis matrices $Q$ where the elements $q_{i,j}$ capture a belief in people transitioning to state (tract) $j$ when currently located at state (tract) $i$ (see Section~\ref{methodology}). 

Our hypotheses are mostly based on existing theories. In this regard, the most prevailing human mobility model is the~\emph{gravitational law}~\cite{zipf1946p1}, which explains mobility by an attraction force between two places that is determined by some weight for each place (e.g., density of venues/points of interest) and the inverted shortest distance between them. 
However, due to some limitations found in this model~\cite{simini2012universal}, many other solutions have emerged to circumvent such problems. 
For example, the~\emph{rank model}~\cite{noulas2012tale} has been proposed as a variation of the radiation model, which indicates that the number of people traveling to a given location is inversely proportional to the number of places surrounding the source location.
Similarly, 
the so called~\emph{intervening opportunities}~\cite{stouffer1940intervening}, 
additionally includes the number of opportunities at a given distance. 
\emph{Cosine similarity} can also be used under the assumption that people prefer to visit places which are similar (according to some metric) to the departure place.
Table~\ref{universal-hyp} summarizes $10$ universal theories considered in this work.

For practical applications these theories require additional data, e.g., to determine the weight of places for the gravitational law.
In our attempt to explain mobility we also utilize online data from Foursquare venues and census data.
We categorize our hypotheses into three types: \emph{Distance-based}, \emph{Foursquare} and \emph{Census} hypotheses, see Table~\ref{tab:hyp-inf}.
Additionally, we use the uniform hypothesis as a baseline to express the belief that all tracts are equally likely for the next stop of a taxi.
For all hypotheses, we set the diagonal of $Q$ to $0$ to avoid self-loop transitions---accounting for only $1.5\%$ of all taxi rides---which (in this work) do not contribute on mobility (i.e., taxi rides that start and end in the same tract).

\para{Distance-based hypotheses.}
Based on the \emph{geographic distance}~\cite{ivis2006calculating}, we can construct very simple and intuitive hypotheses: the \emph{proximity hypothesis} and the \emph{centroid hypothesis}.
The proximity hypothesis assumes that places nearby the current location are more likely to be visited next, the centroid hypothesis suggests that locations near to the city center (specified by fixed geographic coordinates, see Section~\ref{sec:datasets}) are more likely for the next stop.
According to these hypotheses, the location of the next visited place follows a two-dimensional~\emph{Gaussian distribution} that is centered at the current location or the city center respectively, see~\cite{becker2015} for a more detailed description.
For the parametrization of the distribution, we included several variations of the hypotheses with different values for the standard deviation $\sigma$, i.e., $\sigma \in \{0.01, 0.5, 1.0, 2.0, 3.0, 4.0, 5.0\}km$.

\begin{table*}[t!]
	\begin{adjustwidth}{-.01in}{-.01in}
		\centering
		\caption{\textbf{Tract properties.} These three categories contain all tract indicators, i.e., statistics about tracts, used in combination with universal theories to construct hypotheses.
		For instance, the hypothesis \emph{church} expresses that the probability of visiting a certain location is proportional to its number of churches (using for example the density theory).
		}
		\label{tab:hyp-inf}
		\begin{tabular}{|c|l|} \hline 
			\textbf{Category} & \textbf{Properties per tract} \\ \hline
			\multirow{2}{*}{Distance-based} & \multirow{2}{*}{\parbox{15cm}{Fixed points of interest: \emph{Geographical Center}, \emph{Flatiron Building}, \emph{Times Square}.}} \\ 
			& \\ \cline{1-2}
			\multirow{5}{*}{Foursquare} & \multirow{5}{*}{\parbox{15cm}{Number of venues: \emph{Arts \& Entertainment, Education (i.e., colleges, universities, elementary schools and high schools), Food, Nightlife Spot, Outdoors \& Recreation, Work (i.e., auditoriums, buildings, convention centers, event space, factories, government buildings, libraries, medical centers, military base, non-profit, office, post office, prison, radio station, recruiting agency, TV station, and ware house), Residence, Shop \& Service, Travel \& Transport, and Church}.}} \\ 
			& \\
			& \\
			& \\
			& \\ \cline{1-2}
			\multirow{5}{*}{Census} & \multirow{5}{*}{\parbox{15cm}{\emph{Population size}, and \emph{Tract area}. Percentage of: \emph{White people, Black people, People in labor force, Unemployed people, People below poverty level, People above poverty level}. Number of places: \emph{Libraries, Art Galleries, Theaters, Museums, WiFi Hotspots, and Places of Interest}. Moreover, the occupied area of: \emph{Residential Zoning, Commercial Zoning, Manufacturing Zoning, Park properties, Historic Districts and Empower zones}.}} \\ 
			& \\
			& \\
			& \\
			& \\ \cline{1-2}
		\end{tabular}
	\end{adjustwidth}
	
\end{table*}

\para{Foursquare hypotheses.}
To build more informed hypotheses, we leverage Foursquare venues by measuring the \emph{density} of a place, which consists of counting the number of all venues (regardless of their categories) in a given location. Similarly, the number of check-ins can be used to measure the \emph{popularity} of a given place.  
These can be combined with the geographical distance in various ways according to mobility theories, see theories No. $6, 7, 8$ and $9$ in Table~\ref{universal-hyp}.
Additionally, we group places according to their category (e.g., residence, shop or church) based on venues of a single type. 
In other words, we use categories as filters of the available venues. Thus, every Foursquare category induces a subset of all venues per state (i.e., tract). Table~\ref{tab:hyp-inf} shows all $10$ categories included in this study.
To avoid an abundant amount of hypotheses, we only use the gravitational target theory in combination with the category-based hypotheses.
Furthermore, we construct a similarity-based hypothesis that suggests that transitions are more likely between two states that have a similar category distribution of venues based on \emph{Cosine similarity}, see Table~\ref{universal-hyp}.

\para{Census hypotheses.}
Similar to the Foursquare hypotheses, the Census hypotheses add relevant information to every state (i.e., tract). This information on demographics (e.g., \% of white people), land-use (e.g., residential zoning) or socio-economics (e.g., \# of people above poverty level) can be used instead of the number of venues (i.e., density) of a given state. Table~\ref{tab:hyp-inf} shows all $20$ indicators used to formulate this kind of hypotheses. Similarity measures (i.e., cosine similarity) were obtained under three different categories: Race Group (i.e., white, black, american indian, asian, hawaiian and other pacific islander, other race, two races), Poverty Level (i.e., below, above) and Employment status (i.e., employed, unemployed, in labor force).

Overall we defined $70$ hypotheses: (a) $1$ uniform, (b) $29$ distance-based (i.e., geographical distance, proximity and 3 centroids times $7$; for each value of $\sigma$), (c) $17$ from Foursquare (i.e., all venues integrated in formulas $4$ to $9$ from Table~\ref{universal-hyp}, gravitational target for each Foursquare category and $1$ similarity), and (d) $23$ from Census data (i.e., $20$ from the gravitational target with each census indicator and $3$ from similarities.).

\begin{figure*}[t!]
	\centering
	\begin{minipage}{\textwidth} 
		\begin{minipage}{0.45\textwidth}
			\centering
			\subfloat[Temporal distribution of clusters]{
				\includegraphics[width=.9\linewidth]{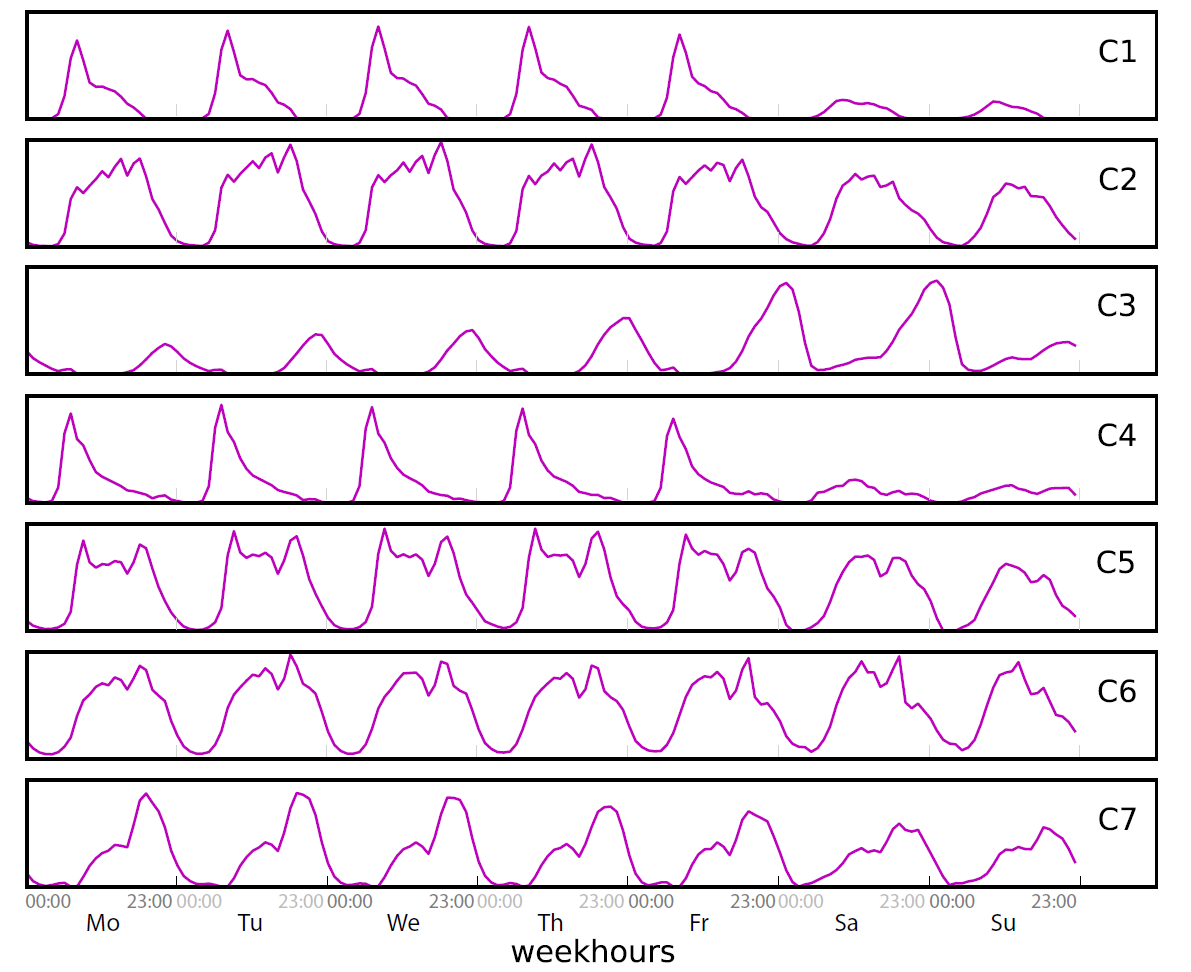}
				\label{fig:ntf-time}
			}
		\end{minipage}
		\begin{minipage}{0.55\textwidth}
			\begin{minipage}{0.31\textwidth}
				\centering 
				\subfloat[Departures C1]{
					\includegraphics[width=0.9\linewidth]{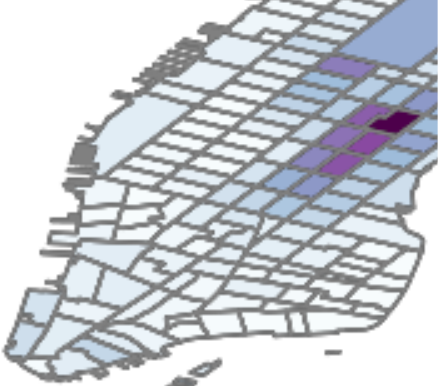} 
					\label{fig:ntf-c1d} 
				}
			\end{minipage}
			\begin{minipage}{0.31\textwidth}
				\centering 
				\subfloat[Departures C2]{
					\includegraphics[width=0.9\linewidth]{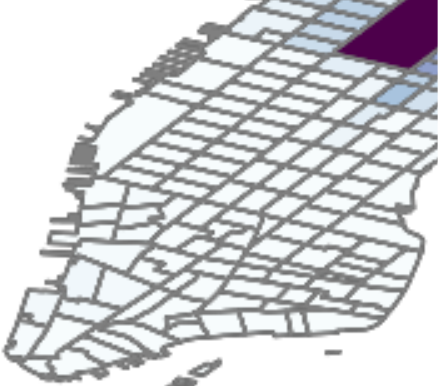} 
					\label{fig:ntf-c2d} 
				}
			\end{minipage}
			\begin{minipage}{0.31\textwidth}
				\centering 
				\subfloat[Departures C3]{
					\includegraphics[width=0.9\linewidth]{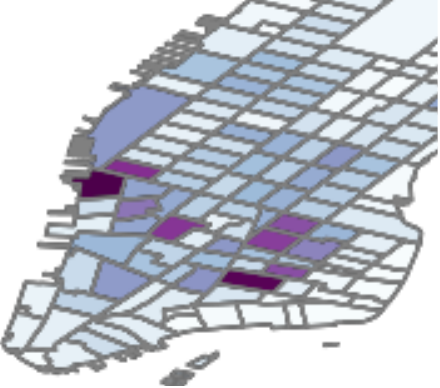}
					\label{fig:ntf-c3d} 
				}
			\end{minipage}\par
			\vspace{1.8em}
			\begin{minipage}{0.31\textwidth}
				\centering 
				\subfloat[Arrivals C1]{
					\includegraphics[width=0.9\linewidth]{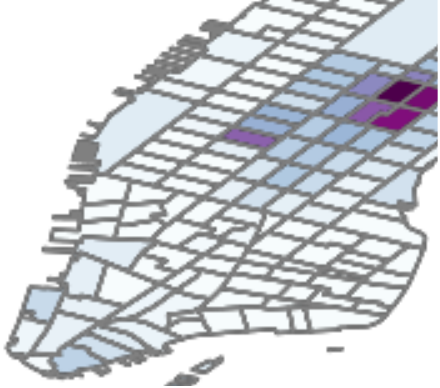} 
					\label{fig:ntf-c1a} 
				}
			\end{minipage}
			\begin{minipage}{0.31\textwidth}
				\centering 
				\subfloat[Arrivals C2]{
					\includegraphics[width=0.9\linewidth]{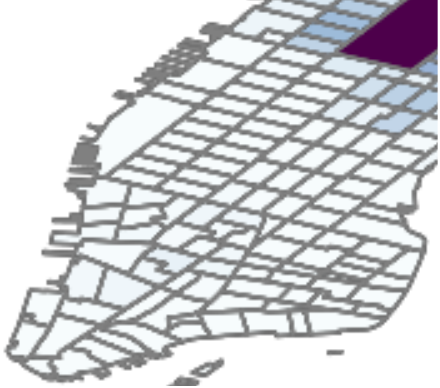} 
					\label{fig:ntf-c2a} 
				}
			\end{minipage}
			\begin{minipage}{0.31\textwidth}
				\centering 
				\subfloat[Arrivals C3]{
					\includegraphics[width=0.9\linewidth]{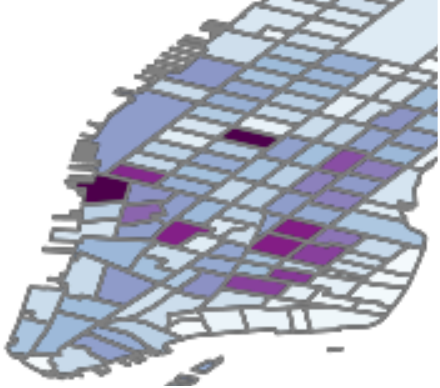} 
					\label{fig:ntf-c3a} 
				}
			\end{minipage}
		\end{minipage}
		
	\end{minipage} 
	\caption{\textbf{Spatio-Temporal behavior obtained by tensor factorization.} This figure illustrates the results obtained when applying NTF on a three-way tensor. (a) Each row represents a behavioral component respect to pickup time (hour of week). Maps shown at the right of this figure show low and midtown Manhattan, the color of each tract determines how representative that location is in that cluster; the darker, the more dominant. (b,e) cluster $C1$ stands for taxis rides at $9am$ around the south-east of Manhattan; (c,f) Cluster $C2$ contains all \emph{evening} short taxi rides at $6pm$ around the center of Manhattan; and (d,g) Cluster $C3$ represents all weekend night taxi rides around the lower and middle part of Manhattan.}
	\label{fig:results}
\end{figure*}

\section{Experiments}
\label{results}
This section reports on experimental results obtained by applying the presented methods to the Manhattan taxi rides dataset.

\subsection {Configuration}
As described in Section~\ref{methodology}, we started by using NTF in order to identify different clusters in the taxi data. 
We worked with a three-way tensor whose dimensions represent the pickup hour of week, the tract of the pickup of the taxi ride and the tract of the drop-off of the taxi ride.
For the hour of week, the first state corresponds to the time from 12:00 a.m. to 12:59 a.m. on Monday, the last of the 168 states to the time from 11:00 p.m. to 11:59 p.m. on Sunday. We did not include time of the drop-off as an extra dimension because most of the rides last less than an hour.
Since there are $288$ different tracts, the tensor for the decomposition has size $168 \times 288 \times 288$.
After experimenting with different parameters, we
set the number of clusters to $r=7$ to cluster the data into seven different groups, as this number subjectively best captured all behavioral components.

\subsection{NTF: Mobility patterns}
Since our data tensor has three dimensions, the decomposition returned three different components which determine the scale of mobility flow in each dimension for every cluster: time (hour of week) and space (departure and arrival tracts). Thus, individual clusters represent groups of taxi rides in different places at different periods of time in a weekday-hour scale.

The time component is shown in Figure~\ref{fig:ntf-time}, whereas location components for pickups (departures) and drop-offs (arrivals) are shown on the right-hand side of Figure~\ref{fig:results}. Due to space limitations we show only the first three clusters.
From Figure~\ref{fig:ntf-time}, it can be observed that all clusters show strong daily regularities, which can be assumed as daily routines in human mobility behavior. Clusters $C1$, $C2$, $C4$, $C5$ and $C7$ capture all behaviors on workdays (almost all peaks are within the first five days of the week), whereas cluster $C3$ is strongly dominated by weekends, specially by Friday and Saturday nights. Cluster $C6$ on the other hand, shows a more periodical behavior across the entire week, however its peaks are around $6pm$ from Monday to Saturday and $2pm$ on Saturdays and Sundays.

The location components shown from Figure~\ref{fig:ntf-c1d} to Figure~\ref{fig:ntf-c3a}, together with their respective time periods, provide us initial context about \emph{when} and \emph{where} people move within the city. For instance, cluster $C1$ represents morning taxi rides around $9am$ (see Fig.~\ref{fig:ntf-time}) which go to the south-east of Manhattan (see Fig.~\ref{fig:ntf-c1a}). Cluster $C2$ on the other hand, concentrates its taxis rides in the evening around $6pm$ near Central Park (see Fig.~\ref{fig:ntf-c2a}). Finally, cluster $C3$ includes all taxi rides on Friday and Saturday around $1am$ in the lower part of Manhattan (see Fig.~\ref{fig:ntf-c3a}).

The following section presents 
characterizations for such behaviors by comparing a list of $70$ hypotheses using HypTrails.

\subsection{HypTrails: Ranking of hypotheses}
Since NTF only assigns a weight to every instance of each factor (i.e., does not explicitly partition the transition input), we first need to identify all transitions for each cluster in order to run HypTrails. 
Generally \cite{peng2012collective,takeuchi2013non}, interpreting clusters returned by NTF requires to extract the top-$N$ most representative states from each factor to determine the instances that are part of every cluster. 
In our setting, we extracted the top-$10$  pickup weekday-hours and drop-off tracts, and query all taxi rides fulfilling these conditions. 
Note that we did not include the top pickup tracts, because we are interested on places where people go to, rather than places where they come from. 
We then applied HypTrails and computed the rankings for the weighting factor $k=10$---see Section~\ref{discussion} for a discussion. 

Exemplary results of the characterization step for a hand selected subset of hypotheses are displayed in Table~\ref{tbl:res-hyp}.
It shows for each hypothesis the respective rank in the cluster; lower numbers imply a higher rank and therefore a better explanation.
We show the results of the \emph{Foursquare} and \emph{Census} hypotheses only applied to the \emph{gravitational target} formula (see Section~\ref{hypotheses}). Gravitational laws (i.e., gravitational target and gravitational mass) in all cases perform the same due to the normalizations applied in the HypTrails approach. Thus, from now onwards they will be referred to as \emph{gravitational}.
Thus, for instance, the hypothesis \emph{Gravitational (\% White people)} expresses a belief in people going to \emph{nearby} tracts with a \emph{high percentage of white people} living there.
In the column \emph{Overall} of Table~\ref{tbl:res-hyp}, we show the ranking of hypotheses evaluated over the whole dataset, to compare with the results obtained for the individual clusters. 
The uniform hypothesis is a baseline which allows us to verify whether a hypothesis can be a good explanation of human mobility or not. 
Green cells in the table indicate that a hypothesis performed better than the uniform hypothesis in that cluster.

To characterize the different patterns in human mobility (i.e., the cluster results of the NTF step), we inspect the obtained rankings for the different clusters.
In particular, we are interested in which hypotheses perform exceptionally well (i.e., have a top rank indicated by a small number) in the cluster both on an absolute scale and in comparison to the ranking obtained from the overall dataset.
For example, consider Cluster $C3$ that captures time frames at the weekend nights.
For this cluster, we can observe very high ranks for the gravitational hypotheses \emph{Party}, \emph{Popularity} (check-ins), \emph{Food} and \emph{Recreation}.

In summary, clusters $C1, C2$ and $C3$ (shown in Figure~\ref{fig:results}) can be characterized as follows. Cluster $C1$ predominantly represents taxi rides around $9am$ on workdays. People in this cluster prefer to go to nearby tracts containing popular places such as work places and restaurants near Times Square in a radio of $0.5km$. Cluster $C2$ groups taxi rides going to big tracts containing art galleries, museums and parks. Transitions in this group usually leave tracts with a small number of venues on workdays around $6pm$. 
Finally, cluster $C3$ identifies all taxi rides on Fridays and Saturdays night around $1am$. People in this cluster usually go to nearby tracts containing very popular places such as nightlife spots and restaurants. Below, we discuss the characteristic properties of all clusters summarized by the types of hypotheses.

\begin{table*}[t!] 
	\centering
	\caption{\textbf{Ranking of Hypotheses}. This table shows the ranking of $23$ out of $70$ hypotheses evaluated with HypTrails over $3$ different groups. \emph{Overall} represents all $143M$ taxi rides in Manhattan $2013$, clusters \emph{$C_i$} are clusters identified by NTF. Numeric cells represent the ranks of the hypotheses  in respective clusters. For the distance-based hypotheses, we only show results for the best parameter of the standard deviation $\sigma$ (parameter in parentheses). Green cells highlight all hypotheses that outperform the uniform hypothesis.}
	
	\renewcommand{\arraystretch}{1.00}
	\label{tbl:res-hyp}
	\footnotesize
	\begin{tabular}{|l|c|c|c|c|c|c|c|c|}
		\hline
		\multicolumn{1}{|c|}{} & \textbf{Overall} & \textbf{C1} & \textbf{C2} & \textbf{C3} & \textbf{C4} & \textbf{C5} & \textbf{C6} & \textbf{C7} \\
		\multicolumn{1}{|c|}{\multirow{-2}{*}{\textbf{HYPOTHESES}}} & \emph{2013} & \emph{\begin{tabular}[c]{@{}c@{}}Workdays\\ 9am\end{tabular}} & \emph{\begin{tabular}[c]{@{}c@{}}Workdays\\ 6pm\end{tabular}} & \emph{\begin{tabular}[c]{@{}c@{}}Weekends\\ 1am\end{tabular}} & \emph{\begin{tabular}[c]{@{}c@{}}Workdays\\ 7am\end{tabular}} & \emph{\begin{tabular}[c]{@{}c@{}}Workdays\\ 9am, 6pm\end{tabular}} & \emph{\begin{tabular}[c]{@{}c@{}}Mo-Sa 6pm\\ Sa-Su 2pm\end{tabular}} & \emph{\begin{tabular}[c]{@{}c@{}}Workdays\\ 6pm\end{tabular}} \\ \hline
		\hline
		\multicolumn{9}{|c|}{\textbf{Baseline}} \\ \hline
		Uniform & 42 & 56 & 56 & 56 & 56 & 55 & 62 & 59 \\ \hline
		\multicolumn{9}{|c|}{\textbf{Distance-based ($\sigma$)}} \\ \hline
		Proximity & \cellcolor[HTML]{C9F4C8}14 (3.0) & \cellcolor[HTML]{C9F4C8}7 (1.0) & \cellcolor[HTML]{C9F4C8}2 (0.5) & \cellcolor[HTML]{C9F4C8}14 (1.0) & \cellcolor[HTML]{C9F4C8}10 (1.0) & \cellcolor[HTML]{C9F4C8}19 (1.0) & \cellcolor[HTML]{C9F4C8}10 (0.01) & \cellcolor[HTML]{C9F4C8}13 (1.0) \\ \hline
		Centroid (Geographical Center) & \cellcolor[HTML]{C9F4C8}38 (5.0) & \cellcolor[HTML]{C9F4C8}50 (5.0) & \cellcolor[HTML]{C9F4C8}25 (1.0) & 58 (5.0) & \cellcolor[HTML]{C9F4C8}52 (5.0) & 58 (5.0) & \cellcolor[HTML]{C9F4C8}51 (3.0) & \cellcolor[HTML]{C9F4C8}51 (5.0) \\ \hline
		Centroid (Flatiron Building) & \cellcolor[HTML]{C9F4C8}29 (5.0) & \cellcolor[HTML]{C9F4C8}32 (2.0) & \cellcolor[HTML]{C9F4C8}51 (5.0) & \cellcolor[HTML]{C9F4C8}17 (1.0) & \cellcolor[HTML]{C9F4C8}2 (0.01) & \cellcolor[HTML]{C9F4C8}4 (0.5) & \cellcolor[HTML]{C9F4C8}44 (3.0) & \cellcolor[HTML]{C9F4C8}20 (0.5) \\ \hline
		Centroid (Times Square) & \cellcolor[HTML]{C9F4C8}22 (3.0) & \cellcolor[HTML]{C9F4C8}1 (0.5) & \cellcolor[HTML]{C9F4C8}43 (3.0) & \cellcolor[HTML]{C9F4C8}46 (3.0) & \cellcolor[HTML]{C9F4C8}1 (0.5) & \cellcolor[HTML]{C9F4C8}43 (2.0) & \cellcolor[HTML]{C9F4C8}2 (0.01) & \cellcolor[HTML]{C9F4C8}1 (0.01) \\ \hline
		\multicolumn{9}{|c|}{\textbf{Foursquare}} \\ \hline
		Gravitational (All venues) & \cellcolor[HTML]{C9F4C8}1 & \cellcolor[HTML]{C9F4C8}12 & \cellcolor[HTML]{C9F4C8}14 & \cellcolor[HTML]{C9F4C8}10 & \cellcolor[HTML]{C9F4C8}14 & \cellcolor[HTML]{C9F4C8}10 & \cellcolor[HTML]{C9F4C8}14 & \cellcolor[HTML]{C9F4C8}9 \\ \hline
		Gravitational (Check-ins) & \cellcolor[HTML]{C9F4C8}9 & \cellcolor[HTML]{C9F4C8}3 & \cellcolor[HTML]{C9F4C8}30 & \cellcolor[HTML]{C9F4C8}2 & \cellcolor[HTML]{C9F4C8}11 & \cellcolor[HTML]{C9F4C8}5 & \cellcolor[HTML]{C9F4C8}4 & \cellcolor[HTML]{C9F4C8}3 \\ \hline
		Gravitational (Work) & \cellcolor[HTML]{C9F4C8}2 & \cellcolor[HTML]{C9F4C8}5 & \cellcolor[HTML]{C9F4C8}12 & \cellcolor[HTML]{C9F4C8}24 & \cellcolor[HTML]{C9F4C8}8 & \cellcolor[HTML]{C9F4C8}6 & \cellcolor[HTML]{C9F4C8}13 & \cellcolor[HTML]{C9F4C8}11 \\ \hline
		Gravitational (Food) & \cellcolor[HTML]{C9F4C8}5 & \cellcolor[HTML]{C9F4C8}4 & \cellcolor[HTML]{C9F4C8}31 & \cellcolor[HTML]{C9F4C8}4 & \cellcolor[HTML]{C9F4C8}12 & \cellcolor[HTML]{C9F4C8}15 & \cellcolor[HTML]{C9F4C8}11 & \cellcolor[HTML]{C9F4C8}4 \\ \hline
		Gravitational (Party) & \cellcolor[HTML]{C9F4C8}7 & \cellcolor[HTML]{C9F4C8}17 & \cellcolor[HTML]{C9F4C8}37 & \cellcolor[HTML]{C9F4C8}1 & \cellcolor[HTML]{C9F4C8}19 & \cellcolor[HTML]{C9F4C8}9 & \cellcolor[HTML]{C9F4C8}20 & \cellcolor[HTML]{C9F4C8}5 \\ \hline
		Gravitational (Recreation) & \cellcolor[HTML]{C9F4C8}15 & \cellcolor[HTML]{C9F4C8}21 & \cellcolor[HTML]{C9F4C8}10 & \cellcolor[HTML]{C9F4C8}9 & \cellcolor[HTML]{C9F4C8}17 & \cellcolor[HTML]{C9F4C8}13 & \cellcolor[HTML]{C9F4C8}7 & \cellcolor[HTML]{C9F4C8}33 \\ \hline
		Venue Similarity & \cellcolor[HTML]{C9F4C8}39 & \cellcolor[HTML]{C9F4C8}53 & \cellcolor[HTML]{C9F4C8}53 & \cellcolor[HTML]{C9F4C8}53 & \cellcolor[HTML]{C9F4C8}53 & \cellcolor[HTML]{C9F4C8}52 & \cellcolor[HTML]{C9F4C8}58 & \cellcolor[HTML]{C9F4C8}53 \\ \hline
		\multicolumn{9}{|c|}{\textbf{Census}} \\ \hline
		Gravitational (Population) & \cellcolor[HTML]{C9F4C8}21 & 61 & \cellcolor[HTML]{C9F4C8}28 & \cellcolor[HTML]{C9F4C8}20 & 59 & \cellcolor[HTML]{C9F4C8}46 & \cellcolor[HTML]{C9F4C8}42 & \cellcolor[HTML]{C9F4C8}25 \\ \hline
		Gravitational (Tract Area) & \cellcolor[HTML]{C9F4C8}23 & \cellcolor[HTML]{C9F4C8}34 & \cellcolor[HTML]{C9F4C8}8 & \cellcolor[HTML]{C9F4C8}26 & \cellcolor[HTML]{C9F4C8}24 & \cellcolor[HTML]{C9F4C8}20 & \cellcolor[HTML]{C9F4C8}24 & \cellcolor[HTML]{C9F4C8}38 \\ \hline
		Gravitational (\%White people) & \cellcolor[HTML]{C9F4C8}6 & \cellcolor[HTML]{C9F4C8}24 & \cellcolor[HTML]{C9F4C8}13 & \cellcolor[HTML]{C9F4C8}28 & \cellcolor[HTML]{C9F4C8}28 & \cellcolor[HTML]{C9F4C8}27 & \cellcolor[HTML]{C9F4C8}35 & \cellcolor[HTML]{C9F4C8}27 \\ \hline
		Gravitational (Residential zoning) & 50 & 65 & \cellcolor[HTML]{C9F4C8}19 & \cellcolor[HTML]{C9F4C8}35 & 65 & 61 & 67 & \cellcolor[HTML]{C9F4C8}49 \\ \hline
		Gravitational (Commercial zoning) & \cellcolor[HTML]{C9F4C8}13 & \cellcolor[HTML]{C9F4C8}8 & \cellcolor[HTML]{C9F4C8}32 & \cellcolor[HTML]{C9F4C8}22 & \cellcolor[HTML]{C9F4C8}9 & \cellcolor[HTML]{C9F4C8}24 & \cellcolor[HTML]{C9F4C8}19 & \cellcolor[HTML]{C9F4C8}15 \\ \hline
		Gravitational (Art Galleries) & 46 & \cellcolor[HTML]{C9F4C8}23 & \cellcolor[HTML]{C9F4C8}1 & \cellcolor[HTML]{C9F4C8}38 & \cellcolor[HTML]{C9F4C8}5 & \cellcolor[HTML]{C9F4C8}2 & \cellcolor[HTML]{C9F4C8}52 & \cellcolor[HTML]{C9F4C8}54 \\ \hline
		Gravitational (Museums) & 54 & \cellcolor[HTML]{C9F4C8}13 & \cellcolor[HTML]{C9F4C8}3 & \cellcolor[HTML]{C9F4C8}40 & \cellcolor[HTML]{C9F4C8}6 & \cellcolor[HTML]{C9F4C8}7 & \cellcolor[HTML]{C9F4C8}26 & \cellcolor[HTML]{C9F4C8}58 \\ \hline
		Gravitational (Parks) & 63 & 62 & \cellcolor[HTML]{C9F4C8}4 & \cellcolor[HTML]{C9F4C8}44 & 63 & 59 & \cellcolor[HTML]{C9F4C8}6 & 64 \\ \hline
		Race Similarity & \cellcolor[HTML]{C9F4C8}32 & \cellcolor[HTML]{C9F4C8}48 & \cellcolor[HTML]{C9F4C8}50 & \cellcolor[HTML]{C9F4C8}52 & \cellcolor[HTML]{C9F4C8}50 & \cellcolor[HTML]{C9F4C8}50 & \cellcolor[HTML]{C9F4C8}59 & \cellcolor[HTML]{C9F4C8}50 \\ \hline
		Poverty Similarity & \cellcolor[HTML]{C9F4C8}37 & \cellcolor[HTML]{C9F4C8}55 & \cellcolor[HTML]{C9F4C8}52 & \cellcolor[HTML]{C9F4C8}54 & \cellcolor[HTML]{C9F4C8}55 & \cellcolor[HTML]{C9F4C8}53 & \cellcolor[HTML]{C9F4C8}61 & \cellcolor[HTML]{C9F4C8}56 \\ \hline
		Employment Similarity & \cellcolor[HTML]{C9F4C8}40 & 57 & \cellcolor[HTML]{C9F4C8}54 & \cellcolor[HTML]{C9F4C8}55 & 58 & \cellcolor[HTML]{C9F4C8}54 & \cellcolor[HTML]{C9F4C8}60 & \cellcolor[HTML]{C9F4C8}58 \\ \hline
	\end{tabular}
\end{table*}

\para{Distance-based.}
As mentioned in Section~\ref{universal-hyp}, these hypotheses require the standard deviation ($\sigma$) of a two-dimensional Gaussian distribution. In Table~\ref{tbl:res-hyp}, we show the best result for parametrized hypotheses and their respective value of $\sigma$ in parenthesis.
In the overall data as well as clusters $C2$ and $C3$, taxi rides are more likely to visit proximate places in a radio of $3km$, $0.5km$ and $1km$ respectively. Clusters $C1, C4, C6$ and $C7$ show preference on visiting the surroundings of Times Square in a radio of $0.5km$, $0.5km$, $0.01Km$ and $0.01km$ respectively. Finally, taxi rides in cluster $C5$ tend to visit places near the Flatiron building in a radio of $0.5km$.

\para{Foursquare.}
Taxi rides in the overall dataset are more likely to visit very dense areas containing work places, restaurants, discos and bars. Similarly, clusters $C1, C4$ and $C5$ which happen to be morning rides around $7-9am$, prefer to go to tracts containing work places such as buildings and offices. Taxi rides in cluster $C6$ go to tracts dominated by recreation places such as parks and other outdoors in the evenings around $6pm$.
Cluster $C3$ can be best characterized by the party hypothesis which means that taxi rides tend to visit tracts containing nightlife spots (on weekend nights---inferred with NTF). Likewise, in cluster $C7$ people tend to visit tracts containing popular places such as restaurants and nightlife spots. 
Note that all hypotheses in this group perform better than the uniform hypothesis in all clusters, demonstrating the overall high explanatory power of such data sources with respect to human mobility.

\para{Census.}
From the overall data, we can infer that users tend to visit close by tracts with high percentage of white people living in them. 
We can also observe that in general taxi rides are not going to residential areas but to commercial zones, which is also the case of clusters $C1, C3$ and $C7$, opposite to clusters $C2, C4$ and $C5$ where it is more likely to visit tracts containing art galleries and museums. In cluster $C6$ we can deduce that people tend to visit tracts containing parks rather than residential zones.

\section{Discussion}
\label{discussion}
In this work, we have shown that spatio-temporal dynamics in human behavior can potentially be better explained by considering parts of the data separately. 
We identified clusters of taxi rides and utilized openly available data from the Web to explain them. However, there are some aspects that need to be taken into account for the current approach.

\para{Concentration parameter k.} As discussed in~\cite{singer2015hyptrails}, the HypTrails approach requires setting a parameter \emph{k} to elicit Dirichlet priors from hypotheses. Higher values of $k$ express stronger believes in the respective hypotheses. Technically, larger values of $k$ imply higher values of the hyperparameters (pseudo counts) of the Dirichlet distributions. In our experiments, we tried several values of $k$ from $0$ to $100$; overall very similar results. The reported results in this paper use an intermediate value of $k=10$.

\para{Correlations in the explanations.}
Using HypTrails to explain clusters will not be able to identify \emph{causes} of movement patterns, but only \emph{correlations}.
As an example, for Cluster $C2$ in our case study the \emph{Art Galleries} hypothesis performs best. This of course does not mean that taxi trips in that cluster prominently have art galleries as destinations, but that people go to places that also have nearby art galleries.
In that direction, we also intend to integrate a correlation analysis between the used hypotheses in future work in order to identify explanations that are similar to each other.

\para{Clustering method.}
In this paper, we use HypTrails to explain clusters obtained by Non-negative tensor factorization. 
While NTF is a reliable and established method in this line of research, the clustering approach is exchangeable and could be replaced by any other clustering technique.

\para{State space.} Since our approach requires a discrete state space, it is necessary to aggregate pick-up and drop-off locations of the taxi rides in an area. The choice of these aggregational units (i.e., the states in our state space) can potentially influence the results. This is known in literature as the \emph{Modifiable areal unit problem}~\cite{openshaw1983}.
In this paper, we chose tracts for the level of our analysis as it allowed for the direct integration of information from census data.
Experiments with different state spaces are subjects of future work.

\para{Multiple dimensions.} 
In this paper, we cluster taxi trips with respect to time, pick-up and drop-off location.
However, our approach allows to extend these by additional information, e.g., number of passengers. In this case, a higher dimensional tensor would be used by NTF, but the resulting clusters could also be explained with help of the HypTrails approach. The scale of these dimensions could also suggest more fine-grained mobility behaviors. For instance, in this work we defined the time dimension as all $168$ hours of a week in order to 
distinguish patterns on workdays and weekends.

\para{Normalization.}
The external online data we use for constructing mobility hypotheses might not be evenly distributed across all tracts (e.g., \# of churches vs. \# of restaurants). Therefore the HypTrails approach normalizes each belief matrix to be able to compare different hypotheses to each other.

\section{Related Work}
\label{rel-work}

\para{Human mobility research.}
Human mobility is a phenomenon that has attracted the attention of governments and researchers from different fields. Studying the movement of people (e.g., migration or commuting) from a social science perspective has helped us to understand who, where and why people move~\cite{stouffer1940intervening, brettell2014migration, merriman2012mobility, willis2010introduction} as well as what consequences such movement carries, by means of e.g., demographic, socio-economic and land-use factors. In literature, they are also referred to as activity-based analysis. Natural sciences, on the other hand, have shown us that \textit{universal patterns} exist and are modelled by movement-based techniques which can predict human dynamics~\cite{simini2012universal,noulas2012tale}. To the best of our knowledge, there exists little research on explaining human mobility by combining both movement-based approaches and activity-based analyses. Furthermore, they are not able to capture occasional or seasonal behaviors that occur only at certain periods of time.

\para{Ubiquitous data.}
Due to the lack of open and updated information at global scale (e.g., surveys or census data), and thanks to the rise of ubiquitous technologies such as mobile phone data and GPS, researchers can get access to human trails which facilitates the study of human movements. There exist several studies that have revealed spatio-temporal patterns in different cities based on mobile phone call detail records (CDR)~\cite{jiangactivity, toole2012inferring}, taxi trips~\cite{chen2010study, liu2013revealing, ding2015understanding} and bike rides~\cite{kaltenbrunner2008bicycle, sarkar2015comparing}. The rapid emerge of social networks has also benefited the study of human mobility based on geo-tagged data. For instance, the work by Jurdak et al.~\cite{jurdak2014understanding} studies Twitter as a proxy of human movement, by using universal indicators such as \textit{displacement distribution} and \textit{gyration radius distribution} that measure how far individuals typically move based on geo-located tweets. Similarly, the authors in~\cite{noulas2015topological} proposed a network of places built upon Foursquare's venues and model human mobility by considering temporal and network dynamics inferred from user's check-ins. Gabrielli et al.~\cite{gabrielli2014tweets} proposed a technique to analyze human trajectories of residents and tourists by semantically labeling source and destination spots. Based on time-evolving networks, the work in~\cite{gauvin2015revealing} identifies and ranks collective features for epidemic spread. This is a more intrusive way of tracking human movement, since it requires the use of wearable sensors.

\para{Activity-based human behavior.}
Compared to this paper, other works have identified
 and explained periodical movement-based patterns as activity-based human behaviors. In ~\cite{wu2014intra}, the authors proposed a model to represent the transition probability of travel demands during a time interval and suggested that travel demands can be associated with fixed locations under some circumstances. Jiang et al.~\cite{jiang2012clustering} explained when, where and how individuals interact with places in metropolitan areas based on activity survey data in Chicago. The work shows daily patterns as eigenvectors and employs K-means clustering to identify groups of individuals based on their daily activities on weekdays and weekends. From taxi trips in Shanghai, the work in~\cite{peng2012collective} shows how to detect basis patterns for collective traffic flow and correlates them with trip categories and temporal activities such as commuting to/from work in the mornings and evenings or going out at night. Linear combinations are used to describe macro patterns and non-negative matrix factorization (NMF) for detecting how many different patterns exist in a day.

\para{Applied urban computing.} Mobility has also been studied in many other different urban contexts: Salnikov et al.~\cite{salnikov2015openstreetcab} studied taxi ride dynamics in order to recommend people the cheapest way of commuting based on two different taxi companies. In~\cite{falcone2014place, vaca2015taxonomy}, the authors proposed two different techniques to infer land-use or taxonomy of places based on geo-tagged posts from Twitter and Foursquare respectively. Taxi drivers can also benefit from studying their driving behavior while passive (no passengers)~\cite{ding2015understanding} and by predicting the number of passengers they could potentially get in a certain hotspot in the next time interval~\cite{li2012prediction}. Cities can benefit as well by correlating mobility patterns and energy consumption to provide better sustainable urban forms~\cite{le2012urban}, and by measuring their influence in the world on attracting visitors~\cite{lenormand2015human}, and by quantifying their resilience to extreme events~\cite{donovan2015using}. In~\cite{ferreira2013visual, chua2014flowsampler}, the authors implemented visualization tools to understand urban flows across time and space from taxi rides and social media.

The novelty of our approach relies on three factors: (1) a multidimensional pattern recognition process using NTF~\cite{cichocki2009nonnegative} to identify different mobility behaviors in taxi data, (2) the expansion of the activity-based human mobility behavior into a hypothesis-based schema built upon human beliefs and (3) quantifying the plausibility of beliefs for every mobility behavior using HypTrails~\cite{singer2015hyptrails}---a Bayesian approach for expressing and comparing hypotheses about human trails.\section{Conclusions}
In this paper, we have presented an innovative approach for discovering and characterizing patterns in human mobility behavior. It combines (i) the clustering of transition data utilizing non-negative tensor factorization (NTF) with both time and space dimensions and (ii) characterizing these clusters using the Bayesian HypTrails method. 
In our experiments on taxi data from Manhattan, we were able to identify several interesting facets of human mobility and characterize them using census data and additional information collected from Foursquare. 
As one example, we discovered a group of taxi rides that end at locations with a high density of party venues on weekend nights. The strength of this approach relies on the fact that the interpretation of the clustering results 
can be easily characterized with high level hypotheses using HypTrails. 

Our work extends recent research concerned with a better understanding of human mobility. We have demonstrated that human mobility is not one-dimensional but rather contains different facets including (but not limited to)
time and space. 
Future research can benefit from our methodological and experimental concepts presented in this work. A more fine-grained view on human mobility can also facilitate e.g., city planners, traffic control, location-based recommender systems or event planning.

In the future, we aim to generalize our findings by studying similar data (e.g., bike trips or geo-tagged tweets) available for New York and other cities.
In doing so, we could not only unveil novel general patterns of mobility, but also discover similarities and differences between cities.

\para{Acknowledgments.}
This work was partially funded by DFG German Science Fund research projects ``KonSKOE'' and ``PoSTs II''.

{
\scriptsize
\raggedright
\sloppy
\balance
\bibliographystyle{abbrv}

}
\end{document}